\documentclass[prc,showpacs,showkeys,aps,epsfig]{revtex4}

\usepackage{psfig}

\def\lsim{\raise0.3ex\hbox{$<$\kern-0.75em\raise-1.1ex\hbox{$\sim$}}}

\def\gsim{\raise0.3ex\hbox{$>$\kern-0.75em\raise-1.1ex\hbox{$\sim$}}}

\newcommand{\be}{\begin{equation}}

\newcommand{\ee}{\end{equation}}

\def\beq{\begin{equation}}

\def\eeq{\end{equation}}

\def\beqa{\begin{eqnarray}}

\def\eeqa{\end{eqnarray}}

\def\gappeq{\mathrel{\rlap {\raise.5ex\hbox{$>$}}

{\lower.5ex\hbox{$\sim$}}}}

\def\lappeq{\mathrel{\rlap{\raise.5ex\hbox{$<$}}

{\lower.5ex\hbox{$\sim$}}}}

\def\Toprel#1\over#2{\mathrel{\mathop{#2}\limits^{#1}}}

\begin{document}

\title{Constraining the nuclear gluon distribution in $eA$ processes at RHIC}
\author{ E.R. Cazaroto$^1$, F. Carvalho$^1$,  V.P. Gon\c{c}alves$^2$, and  F.S. Navarra$^1$}
\affiliation{$^1$Instituto de F\'{\i}sica, Universidade de S\~{a}o Paulo,
C.P. 66318,  05315-970 S\~{a}o Paulo, SP, Brazil\\
$^2$High and Medium Energy Group (GAME), \\
Instituto de F\'{\i}sica e Matem\'atica,  Universidade
Federal de Pelotas\\
Caixa Postal 354, CEP 96010-900, Pelotas, RS, Brazil\\}

\begin{abstract}

A systematic determination of the gluon distribution is of fundamental
interest in understanding the parton structure of nuclei and the QCD dynamics.
Currently, the behavior of this distribution at small $x$ (high energy) is
completely undefined. In this paper we analyze the possibility of constraining 
the nuclear effects present in $xg^A$ using the inclusive observables which would be measured in the future electron-nucleus collider at RHIC. We demonstrate that the study of nuclear longitudinal and  charm structure functions allows to estimate the magnitude of shadowing and antishadowing effects in the nuclear gluon distribution.

\end{abstract}
\pacs{12.38.-t, 24.85.+p, 25.30.-c}

\keywords{Quantum Chromodynamics, Nuclear Gluon Distribution, Shadowing effect}

\maketitle
\vspace{1cm}

\begin{figure}
\centerline{\psfig{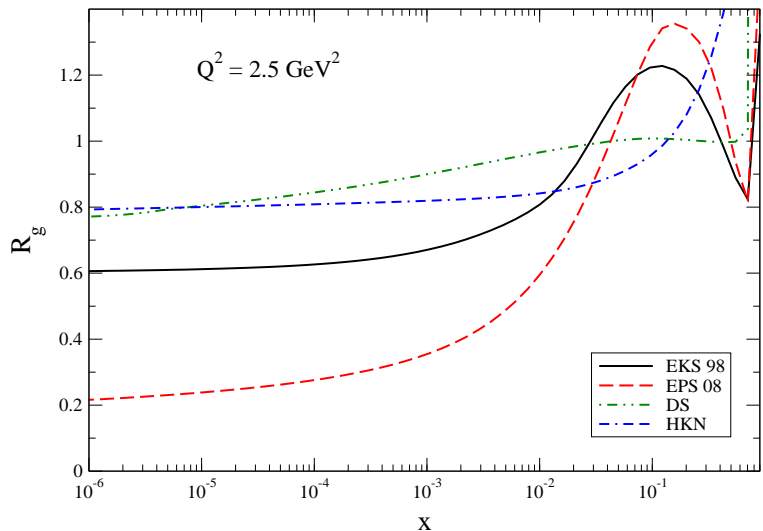}}
\caption{Ratio $R_g = xg^A/A.xg^N$ predicted by the EKS, DS, HKN and EPS 
parameterizations for $A=208$ and $Q^2=10$ GeV$^2$.}
\label{fig1}
\end{figure}

Since the early days of the parton model and of the first deep inelastic scattering  (DIS)  experiments, determining the precise form of the gluon distribution of the nucleon has been   a major goal of high energy hadron physics. Over the last 30 years enormous progress has  been achieved. In particular, data from HERA allowed for a  good determination of the gluon density of the proton.  A much harder task has been to determine the gluon 
distribution of nucleons bound in a nucleus, i.e., the nuclear gluon distribution 
($xg^A (x,Q^2)$). In recent years several experiments have been dedicated to high precision measurements of deep inelastic lepton scattering (DIS) off nuclei. Experiments at CERN and Fermilab focus especially on the region of small values of the Bjorken variable $x = Q^2/2M\nu$, where $Q^2=-q^2$ is the squared four-momentum transfer, $\nu$ the energy transfer and $M$ the nucleon mass.
The data \cite{arneodo}, taken over a wide kinematic range $10^{-5}\,\le\,x\le\,0.1$ and  $0.05\,GeV^2\,\le\,Q^2\le\,100\,GeV^2$, show a systematic reduction of the nuclear structure function $F_2^A(x,Q^2)/A$ with respect to  the free nucleon structure  function $F_2^N(x,Q^2)$. This phenomenon is known as {\it  nuclear shadowing effect} and is associated to  the modification of  the target parton distributions so that $xq^A(x,Q^2) \, < \,  Axq^N(x,Q^2)$, as expected from a superposition of $pp$ interactions (For a review see, e.g. \cite{reviews,armesto}). The modifications depend on the parton
momentum fraction: for momentum fractions $x < 0.1$ (shadowing region)
and $0.3 < x < 0.7$ (EMC region), a
depletion is observed in the nuclear structure functions. These two regions are
bridged by an enhancement known as antishadowing for $0.1 < x < 0.3$.
The experimental data for the nuclear structure function  determine the behavior of the nuclear quark distributions, while the behavior of the  nuclear
gluon distribution is indirectly determined using the momentum sum
rule as a constraint and/or studying the $\log Q^2$ slope of the ratio $F_2^{Sn}/F_2^{C}$
\cite{ehks}.
 Currently, the behavior of $xg^A (x,Q^2)$ at small $x$ (high energy) is completely uncertain as shown in Fig. \ref{fig1}, where we present  the ratio 
$R_g = xg^A/(A.xg^N)$, for $A=208$, predicted by four different groups which realize a global analysis of the nuclear experimental data using the DGLAP evolution equations \cite{dglap} in order to determine the parton densities in nuclei.
In particular, the magnitude of shadowing and the presence or not of the antishadowing effect is completely undefined.

In the last years the analysis of the nuclear effects in deep inelastic scattering (DIS) has been extensively discussed \cite{raju_ea, raju_annual} and motivated by the perspective that in a near future an experimental investigation of the nuclear shadowing at small $x$ and $Q^2 >> 1 \, GeV^2$ using $eA$ scattering could be performed at Brookhaven National Laboratory (eRHIC). It is expected that measurements over the extended $x$ and $Q^2$ ranges, which would become possible at eRHIC, will give more information in order to discriminate between the distinct models of shadowing   and the understanding of the QCD dynamics at small $x$.
This collider is expected to have statistics high  enough to allow for the determination of several inclusive and exclusive observables which are directly dependent on the behavior of the nuclear gluon distribution, as for example, the longitudinal and charm structure functions, the logarithmic slopes with respect to $x$ and $Q^2$, as well as the diffractive leptoproduction  of vector
mesons. In particular, the longitudinal structure function is expected to be measured for the first time in the kinematical regime of small $x$, since the electron - ion collider will be able to vary the energies of both the electron and ion beams.

 In this paper we study the behavior of the nuclear longitudinal structure function  $F_L^A$ and the charm structure function $F_{2 }^{c,A}$ and analyze the possibility to constrain the nuclear effects present in $xg^A$ using these inclusive observables. We estimate the normalized ratios
\beq
R_L(x,Q^2) = \frac{F_L^A(x,Q^2)}{A F_L^p(x,Q^2)} \,\,\,\,\mbox{and}\,\,\,\, R_C(x,Q^2) = \frac{F_{2 }^{c,A}(x,Q^2)}{A F_{2}^{c,p}(x,Q^2)}
\label{rat}
\eeq
considering four distinct parameterizations for the nuclear gluon distributions and compare  their  behavior with those predicted for the  ratio $R_g = xg^A/A.xg^N$. We analyze the similarity between these ratios and  demonstrate that the experimental study of these observables allow to determine the magnitude of shadowing and  antishadowing effects.  We calculate these observables using the Altarelli-Martinelli equation \cite{alta} and the boson-gluon fusion cross section \cite{grv95},
respectively. In other words, we will restrict ourselves to the descriptions which use the DGLAP evolution equations \cite{dglap} to describe the behavior of the nuclear parton
distributions and will assume the validity of the collinear factorization.
It is important to emphasize that the theoretical understanding of
small-$x$ and large $A$ regime of the QCD dynamics  has
progressed in recent years (For recent reviews see, e.g. \cite{cgc}), with the main prediction being  a transition of the
 linear regime described by the DGLAP dynamics to a nonlinear regime where the  
physical process of parton recombination  becomes important in the
parton cascade and the evolution is given by a nonlinear evolution
equation. One of the main motivations for the eRHIC experiment is the study of this new regime, denoted Color Glass Condensate (CGC)  \cite{cgc}. As in Ref. \cite{ea_simone} the inclusive observables at eRHIC were studied using a generalized saturation model, based on the CGC physics, the current study can be considered as complementary to that reference.

Let us start presenting a brief review of  the calculations of the longitudinal and charm structure functions. The longitudinal structure function in deep inelastic scattering is one of the observables from which the gluon distribution can be unfolded. Currently, there is  a expectation for new experimental HERA data for $F_L$ taken with reduced proton energies, which will provide more direct access to the proton gluon distribution and shed light on the QCD dynamics at small-$x$ (See, e.g. Ref. \cite{vicmag_fl}).
Longitudinal photons have zero helicity  and can exist only virtually. In the Quark-Parton Model (QPM), helicity conservation of the electromagnetic vertex yields the Callan-Gross relation, $F_L=0$, for scattering on quarks with spin $1/2$. This does not hold when the quarks acquire transverse momenta from QCD radiation. Instead, QCD yields the Altarelli-Martinelli equation\cite{alta}
\begin{eqnarray}
F_L(x,Q^2) = \frac{\alpha_s(Q^2)}{2\pi}\,x^2\, \int_x^1 \frac{dy}{y^3}[\frac{8}{3}\,F_2(y,Q^2) + 4\,\sum_q e_q^2 (1-\frac{x}{y})yg(y,Q^2)]\,\,,
\label{flalta}
\end{eqnarray}
expliciting the dependence of $F_L$ on the strong coupling constant  and the gluon density. At small $x$ the second term with the gluon distribution is the dominant one. In Ref. \cite{cooper} the authors have suggested  that  expression (\ref{flalta}) can be reasonably approximated  by  $F_L(x,Q^2) \approx 0.3\, \frac{4 \alpha_s}{3 \pi} xg(2.5x,Q^2)$, which  demonstrates  the close relation between the longitudinal structure function and the gluon distribution. Therefore, we expect  
the longitudinal structure function to be sensitive to  nuclear effects. In this paper we  calculate $F_L$ using the Altarelli-Martinelli equation  (\ref{flalta}).

Let us now discuss  charm production and its contribution to the structure function. In the last years, both the H1 and ZEUS collaborations have measured the charm component $F_2^c$ of the structure function at small $x$ and have found it to be a large (approximately $25\%$) fraction of the total \cite{F2CDATA}. 
This is in sharp contrast to what is found at large $x$, where typically 
$F_2^c/F_2\, \approx {\cal{O}}(10^{-2})$. This behavior is directly related to the growth of the gluon distribution at small-$x$. In order to estimate the charm contribution to the structure function we consider the formalism advocated in \cite{grvc} where the charm quark is treated
as a heavy quark and its contribution is given by fixed-order perturbation
theory. This involves the computation of the boson-gluon fusion process. A $c\overline{c}$ pair can be created  by boson-gluon fusion when  the squared  invariant mass  of the hadronic final state is $W^2 \ge 4m_c^2$. Since
$W^2 = \frac{Q^2(1-x)}{x} + M_N^2$, where $M_N$ is the nucleon mass, the charm production  can  occur well below the $Q^2$ threshold, $Q^2 \approx  4m_c^2$, at small $x$. The charm contribution to the proton/nucleus structure function, in leading order (LO), is given by \cite{grv95}
\begin{eqnarray}
\frac{1}{x} F_2^c(x,Q^2,m_c^2) = 2 e_c^2 \frac{\alpha_s(\mu^{\prime 2})}{2\pi}
\int_{ax}^1 \frac{dy}{y}\, C_{g,2}^c(\frac{x}{y},\frac{m_c^2}{Q^2})\,g(y,\mu^{\prime 2})  \,\,,
\label{f2c}
\end{eqnarray}
where $a=1+\frac{4m_c^2}{Q^2}$ and the factorization scale $\mu^{\prime}$ is
assumed $\mu^{\prime 2}=4m_c^2$.  $C_{g,2}^c$ is the coefficient function
given by
\begin{eqnarray}
C_{g,2}^c(z, \frac{m_c^2}{Q^2})  & = & \frac{1}{2} \{ [z^2 + (1-z)^2 +z(1-3z)\frac{4m_c^2}{Q^2} - z^2 \frac{8m_c^4}{Q^4}]
ln \frac{1+\beta}{1-\beta} \nonumber \\ & + & \beta[-1 +8z(1-z) -z(1-z)\frac{4m_c^2}{Q^2}]\}\,\,,
\end{eqnarray}
where $\beta= 1 - \frac{4m_c^2 z}{Q^2 (1-z)}$  is the velocity of one of the charm quarks in the boson-gluon center-of-mass
frame.
Therefore, in leading order, ${\cal{O}}(\alpha_s)$, $F_2^c$ is directly sensitive only to the gluon density via the well-known Bethe-Heitler process
$\gamma^*g \rightarrow c\overline{c}$.
The dominant uncertainty in the QCD calculations arises from the uncertainty
in the  charm quark mass. In this paper we assume $m_c = 1.5\,GeV$.

Finally, let us briefly discuss the distinct parameterizations for the nuclear parton distributions (For details see the  recent review \cite{armesto}).  We will make use of   the existing parameterizations of the nuclear parton distribution functions based on a global fit of the nuclear data using the DGLAP evolution equations. Currently there are four parameterizations, proposed by Eskola, Kolhinen and Salgado \cite{eks98}, by de Florian and Sassot \cite{ds04}, by Hirai, S.~Kumano and T.~H.~Nagai \cite{hkn07} and the very recent one proposed by  K.~J.~Eskola, H.~Paukkunen and C.~A.~Salgado \cite{eps08}. In what follows they will be called EKS, DS, HKN and EPS, respectively.
The basic idea of these approaches is that the experimental results \cite{arneodo} presenting nuclear shadowing effects can be described using the DGLAP evolution equations with adjusted initial parton distributions. Similarly to the  global analyzes of parton distributions in the free proton, they determine the nuclear parton densities at a wide range of $x$ and $Q^2$ through their perturbative DGLAP evolution by using the available experimental data from $lA$ DIS and  $pA$ collisions as a constraint. As pointed out  in Ref. \cite{armesto},  different approaches differ in the 
form of the parameterizations at the initial scale, in the use of different sets of experimental 
data, in the order of the DGLAP evolution, in the different nucleon parton densities used in the 
analysis, in the treatment of isospin effects and in the use of sum rules as additional constraints for the evolution. For instance, the DS and HKN groups provide leading (LO) and next-to-leading order  (NLO) parameterizations, while EKS and EPS perform only a LO QCD global analysis. 
There are noticeable differences between the HKN analysis results and the ones in Ref. \cite{ds04} especially in the strange-quark and gluon modifications. These differences come from various sources. First, the analysed experimental data sets are slightly different. Second, the strange-quark distributions are created by the DGLAP evolution by assuming $s(x)=0$ at the initial $Q^2$ scale, and the charm distributions are neglected in Ref. \cite{ds04}.
These differences lead to  discrepancies among the gluon modifications.
In contrast to the EKS, DS and HKN parameterizations, the EPS one  has included the RHIC data from \cite{BRAHMSdata} in the global fitting procedure. The main assumption is that  these data can be understood with linear evolution.
The inclusion of the high-$p_T$ hadron data from RHIC at forward rapidities  provided important further constraints for the gluon shadowing region. 
By construction, these parameterizations describe the current experimental data. However, the resulting parton distribution sets are very distinct. In particular, the predictions of the different groups for $R_g$ differ largely about the magnitude of the shadowing and the presence or not of the antishadowing. It is associated to the fact that the data included in the global analyses probe the quark distribution, while the gluon  is constrained only 
by the evolution and the momentum sum rule.
 As shown in Fig. \ref{fig1}, while the HKN and DS parameterizations predict a small value of shadowing, the EKS and EPS one predict a large amount, with the distinct predictions differing by a factor 4 at $x = 10^{-5}$. Furthermore, while the DS parameterization does not predict antishadowing and EMC effects in the nuclear gluon distribution, these effects are present in the EKS and EPS parameterizations. In the particular case of the HKN parameterization, it predicts a steep growth of the ratio $R_g$ in the region $x \ge 10^{-1}$. 
It is important to emphasize that the magnitude of shadowing and antishadowing effects in the EKS and EPS parameterizations are  directly related by the momentum sum rule.
The large discrepancies between the predictions of the four parameterization for $xg^A$ in all kinematical $x$ range imply  a large uncertainty in the predictions for  the observables which would be measured  in $pA/AA$ collisions at LHC, for instance. 

\begin{figure}
\centerline{{\psfig{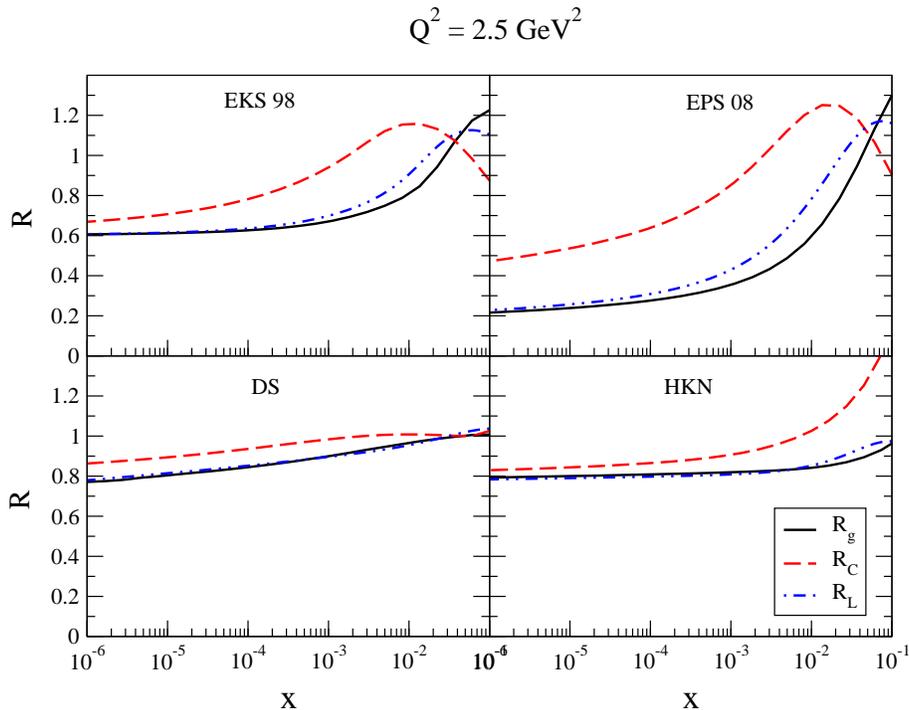}} }
\caption{Ratios $R_g$,   $R_C$  and  $R_L$ for the
 four considered nuclear parameterizations and  $Q^2 = 2.5$ GeV$^2$.}
\label{fig2}
\end{figure}

As mentioned above it is well known that the inclusive observables $F_L$ and $F_2^c$ are strongly dependent on the  gluon distribution. Our goal is to quantify and determine the kinematical region where these observables directly determine the behavior of $R_g$.
In order to obtain model independent conclusions we calculate $R_L$ and $R_C$ using the four parameterizations described above and compare with the corresponding predictions for $R_g$. As the small-$x$ region at eRHIC will be probed at small-$Q^2$ we  concentrate our 
analysis on  two characteristic values of $Q^2$: $Q^2 = 2.5$ GeV$^2$ and $10$  GeV$^2$. Moreover, we only consider $A = 208$, but similar conclusions are obtained for othes values 
of the  atomic number.

\begin{figure}
\centerline{{\psfig{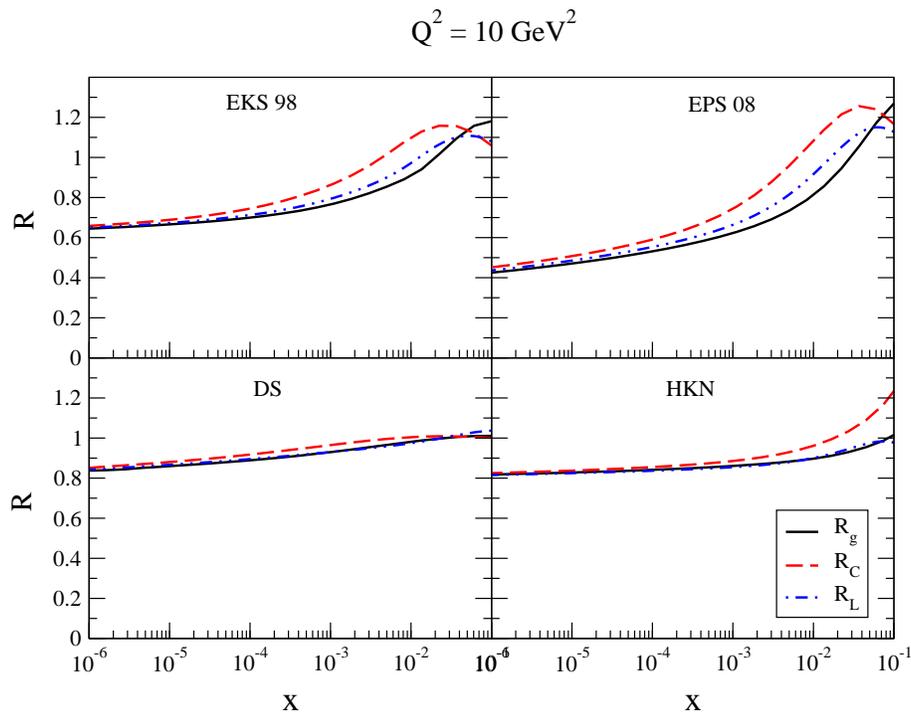}} }
\caption{The same as Fig. \ref{fig2} for  $Q^2 = 10$ GeV$^2$.}
\label{fig3}
\end{figure}

In Figs. \ref{fig2} and \ref{fig3}  we present our results. Firstly, let us  discuss the small-$x$ region, $x \le 10^{-3}$, determined by shadowing effects. We observe  that 
$R_L$ practically coincides with $R_g$ for all parameterizations and for the two values of $Q^2$ considered. This  suggests that  shadowing effects can be easily constrained at eRHIC by  measuring $F_L$. This conclusion is, to a good extent, model independent.
On the other hand, the ratio $R_C$ gives  us an upper bound for the magnitude of the shadowing effects. For example, if it is found that $R_C$ is equal to $\approx 0.6$ at $x = 10^{-4}$ and $Q^2 = 2.5$  GeV$^2$ the  nuclear gluon distributions from 
DS and HKN parameterizations are very large and should be modified. At $Q^2 = 10$ GeV$^2$  
the behavior of $R_C$ is almost identical to $R_g$, which implies that by 
measuring $F_2^c$ at this virtuality we  can also constrain the shadowing effects.
Considering now the kinematical range of $x > 10^{-3}$ we can analyze the correlation between the behavior of $R_L$ and $R_C$ and the antishadowing present or not in the nuclear gluon distribution. Similarly to observed at small values of $x$, the behavior of $R_L$ is very close to the $R_g$ one in the large-$x$ range. In particular, the presence of antishadowing in $xg^A$ directly implies an  enhancement in $F_L^A$. It is almost 10\% smaller in magnitude that the enhancement predicted for $xg^A$ by the EKS and EPS parameterizations. Inversely, if we assume the nonexistence of the antishadowing in the nuclear gluon distribution at $x < 10^{-1}$, as in the DS and HKN parameterizations, no enhancement will be present in $F_L^A$ in this kinematical region. Therefore,  
it suggests that also the antishadowing effects can be easily constrained at eRHIC measuring $F_L$. On the other hand, in this kinematical range the behavior of $R_C$ is distinct of $R_g$ at a same $x$. However, we observe that the behavior of $R_C$ at $x = 10^{-2}$ is directly associated to $R_g$ at $x = 10^{-1}$. In other words, the antishadowing is shifted in $R_C$ by approximately one order of magnitude in $x$. For example, the large growth of $R_g$ predicted by the HKN parameterization at $x \ge 10^{-1}$ shown in Fig. \ref{fig1} implies the  steep behavior of $R_C$ at $x \ge 10^{-2}$   observed in Fig. \ref{fig2}. A similar conclusion can be drawn from Fig. \ref{fig3}. Consequently, by measuring $F_2^c$ it is also
 possible  to constrain the existence and magnitude of the antishadowing effects. 

Some comments are in order here. Firstly, it is important to emphasize that although we have calculated $F_L$ and $F_2^c$ at leading order we expect that the behavior of the ratios $R_L$ and $R_C$ and, consequently, the main conclusions of this paper would not be strongly modified by the NLO corrections. 
Secondly, in our study we only have considered two examples of inclusive observables which would be measured at eRHIC. As demonstrated in Ref. \cite{ea_victor} the study of the logarithmic slope of the nuclear structure function is another important quantity to probe the nuclear effects and the QCD dynamics at small-$x$. Furthermore, the  exclusive production of vector mesons is an important complementary observable to determine the nuclear gluon distribution, since in this case the total cross section is proportional to the square of $xg^A$ (See e.g. Ref. \cite{vicmag_mesons}). Finally, we have disregarded the presence of non-linear effects in the QCD dynamics and used the current parameterizations  based on the DGLAP dynamics, extrapolating them  to lower values of $x$. Consequently, our results can be regarded as conservative and serve as a baseline.  Deviations from this baseline  may indicate the  emergence of the saturation regime of QCD.

Summarizing,  our results indicate that the study of the inclusive observables $F_L$ and $F_2^c$ in $eA$ process at eRHIC is ideal to constrain the nuclear effects present in the nuclear gluon distribution, which, in turn,  is a crucial ingredient to estimate the
cross sections of the processes which will be studied in the future accelerators.
Basically, we see  that  by measuring these observables  we will have a direct access to the nuclear gluon distribution and allow to discriminate between the different parameterizations. We hope that this paper can  motivate a more accurate determination of $F_L$ and $F_2^c$ in the next years.

\begin{acknowledgments}
  This work was  partially financed by the Brazilian funding
agencies CNPq, FAPESP and FAPERGS.
\end{acknowledgments}

\hspace{1.0cm}

\end{document}